\documentclass[10pt, titlepage, twocolumn]{article}

\usepackage{titlesec}
\usepackage{fancyhdr}%
\usepackage{lipsum}
\usepackage{tikz}
\usepackage{siunitx}
\usepackage{pgfplots}
\usepackage{pgfplotstable}
\usepackage{graphicx}
\usepackage{mathtools}
\usepackage[printwatermark]{xwatermark}
\usepackage{xcolor}
\usepackage[labelfont=bf]{caption}
\usepackage[T1]{fontenc}
\usepackage[utf8]{inputenc}
\usepackage{blindtext}
\usepackage{mathrsfs}
\usepackage{gensymb}
\usepackage{float}
\usepackage{algpseudocode,algorithm,algorithmicx}
\usepackage{eso-pic}
\usepackage{amssymb}

\DeclareMathOperator*{\argmax}{argmax}

\graphicspath{ {/} }

\pgfplotsset{compat=newest} %
\usepgfplotslibrary{units} %

\titleformat{\section}
{\normalfont\scshape\centering}{\thesection}{1em}{}

\titleformat{\section}
{\normalfont\Large\bfseries}{\thesection}{0em}{}[{\titlerule[0.8pt]}]

\titlespacing\section{0pt}{10pt plus 4pt minus 2pt}{10pt plus 2pt minus 2pt}

\newcommand{\ii}{\indent\indent}

\newcommand{\drawPlayers}{

		\draw (0,3) circle (.5cm);
		\draw (0,2)--(0,2.5);
		\draw (0,1) -- (0,2);%
		\draw (0,1) -- (1,1);
		\draw (1,1)--(0,0);
		\draw (0,2) -- (.5,1.8) -- (1,1.8);

		\draw (4,3) circle (.5cm);
		\draw (4,2)--(4,2.5);
		\draw (4,1) -- (4,2);%
		\draw (4,1) -- (3,1);
		\draw (3,1)--(4,0);
		\draw (4,2) -- (3.5,1.8) -- (3,1.8);

		\draw (.75,1.70) -- (1.40,1.70);
		\draw (1.40,1.70) -- (1.35, 2.40);
		\draw (2.65,1.70) -- (3.30,1.70);
		\draw (2.65,1.70) -- (2.70, 2.40);

		\draw (.5,1.60) -- (3.5,1.60);
		\draw (2,1.60) -- (2,.20);
		\draw (2,.20)--(1.5,0);
		\draw (2,.20)--(2.5,0);
		\draw (1.3,0)--(1.5,0);
		\draw (2.5,0)--(2.7,0);

		\draw (-0.5,.9)--(.5,.9);
		\draw (-0.5, .9) -- (-.5,0);
		\draw (.5, .9) -- (.5,0);
		\draw (-0.5, .9) -- (-.55,1.2);
		\draw (-.55,1.2)--(-.55,2);

		\draw (3.5,.9)--(4.5,.9);
		\draw (4.5, .9) -- (4.5,0);
		\draw (3.5, .9) -- (3.5,0);
		\draw (4.5, .9) -- (4.55,1.2);
		\draw (4.55,1.2)--(4.55,2);

}

\makeatletter
\renewcommand\thesection{}
\renewcommand\thesubsection{\@arabic\c@section.\@arabic\c@subsection}
\makeatother

\newcommand\BackgroundPicSmall{%
\put(0,365){%
\parbox[b][\paperheight]{\paperwidth}{%
\vfill
\centering
\includegraphics[width=0.05\paperwidth,height=0.05\paperheight,%
keepaspectratio]{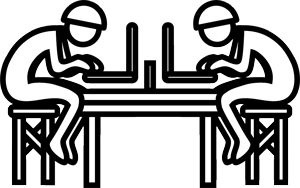}%
\vfill
}}}

\newcommand\BackgroundPicSmallSig{%
\put(180,-150){%
\parbox[b][\paperheight]{\paperwidth}{%
\vfill
\centering
\includegraphics[width=0.15\paperwidth,height=0.15\paperheight,%
keepaspectratio]{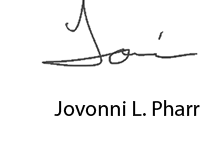}%
\vfill
}}}

\begin{document}

\AddToShipoutPicture{\BackgroundPicSmall}

\begin{titlepage}

\protect\parbox{.9\textwidth}
	{\protect\centering 
		\huge Hacker Combat: A Competitive Sport from Programmatic Dueling \& Cyberwarfare
	}
\begin{center}
{\LARGE
\today}
\end{center}

{
\LARGE
\begin{center}
Department of Computer Information Systems\\
Georgia State University

Atlanta, GA, U.S.A\\
\end{center}
}

\vfill
{
\Large
\begin{center}
Jovonni L. Pharr\\
jpharr2@student.gsu.edu \\
\end{center}
}

{
\Large
\begin{center}
\textbf{Sponsored by}

the Arizona Cyber Warfare Range, Phoenix, AZ, U.S.A
\end{center}
}

\vfill
\section{Abstract}
\noindent
\textit{
	The history of humanhood has included competitive activities of many different forms. Sports have offered many benefits beyond that of entertainment. At the time of this article, there exists not a competitive ecosystem for cyber security beyond that of conventional capture the flag competitions, and the like. This paper introduces a competitive framework with a foundation on computer science, and hacking. This proposed competitive landscape encompasses the ideas underlying information security, software engineering, and cyber warfare. We also demonstrate the opportunity to rank, score, \& categorize actionable skill levels into tiers of capability. Physiological metrics are analyzed from participants during gameplay. These analyses provide support regarding the intricacies required for competitive play, and analysis of play. We use these intricacies to build a case for an organized competitive ecosystem. Using previous player behavior from gameplay, we also demonstrate the generation of an artificial agent purposed with gameplay at a competitive level.
	}
\end{titlepage}

\tableofcontents
\listoffigures
\listofalgorithms

\clearpage

\onecolumn
\section{Manifesto}
It is with great pleasure that this work be publicly shared. At the time of writing, there exists an ever-growing competitive landscape in information security. With cyber warfare being an increasingly pertinent threat to the humanity of civilization, the efficient training of the next generation of cyber warriors becomes ever more important; history has not been kind to the Black Hat. Plotting control over the individuals of whom can liberate an entire generation. These unfortunate state of affairs serve as a guiding beacon throughout this work. This proposal is nation state agnostic, and serves the ethos of the Hacker culture. 

It is understood that this proposal, rightfully belonging to the people, upon being misused, can be detrimental to the purity of the ethos. It is for this reason that the researchers of this project intend to organize this collective effort. All collaborators on this project have exclusively spent volunteer hours to bring this project to completion. It is for this reason that the culture must ensure that no entity, public nor private, succeed in exploiting this framework for purposes that indicate skullduggery.

It is the ethos that widen the window of imagination, and it is the ethos that shall ensure survival of unhindered exploration. The powers that are incentivized to destroy free exploration have not a motivation to empower the ethos with such a framework. This is the reason that this project is \textbf{For Hackers, Built By Hackers}. The supporters of this project are only messengers of the framework. This framework does not invent a new concept, but focuses on disambiguating a competitive structure for cyber warfare, computer science, and information security. In other words, the pieces are already "there", and have been so for quite some time. 

This project does not pretentiously try to convince the reader of a completely unique idea. We do not attempt to take credit for creating this framework. It is fully believed that every practitioner in these fields already perform the behaviors used to partake in the proposed system. This paper is merely a combination of existing techniques, and methodologies.

The instance in which any individual nation state attempts to own this framework, will be a great loss to the entirety of the ethos. However, the capabilities of this type of framework are conducive to militaristic training, but should not be restricted for use by anyone interested. Upon the death, or earthly removal of the project originators, there shall be no bearing on the potential uses of this framework for the remainder of civilization. This proposal is not only created with care, but with passion as well. \textbf{To the ethos, go the freedom to ascend}.

\AddToShipoutPicture*{\BackgroundPicSmallSig}

\clearpage

\twocolumn

\iffalse
\begin{figure}[ht]
	\centering
	%
	\begin{tikzpicture}
		\centering
		
		\drawPlayers

		%
		%
		
	\end{tikzpicture}
	\caption{A setup for head-to-head cyber security combat.}
	\label{normal_case}
\end{figure}
\fi

\begin{figure}[ht]
\centering
	\includegraphics[width=.25\textwidth]{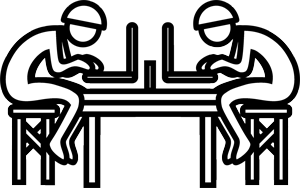}
	\caption{A setup for head-to-head cyber security combat.}
	\label{hackercombatlogo_edit_smallicon}
\end{figure}

\section{Introduction}

Arguably, computer programming is an art form. With every art form, there exists tiers of talent -- novice to expert. This article yields a dueling system built entirely on the methods \& tactics used by computer scientists, and cyber security researchers. Cyber security is an important topic for the safety of the planet, but this paper merely builds on the fundamental philosophy of attack and defense.

At the time of this paper, there exists efforts to introduce Artificial Intelligence to cyber security competitions. These efforts remove the human as the hacker, replaced by an artificial agent. The human plays the role of designing a system capable of autonomous detection, exploitation, and patching of security vulnerabilities. While this is subjectively an interesting field of research, this article focuses on keeping the humanistic actor as the center participant in cyber warfare competition.  

\section{Background}

Today, there exists an ecosystem of capture-the-flag competitions related to the cyber security industry. This ecosystem is constantly growing, and offers many benefits to the field. Traditionally, Black Hat Capture the Flag games (CTF) are held, sometimes referred to as "Red versus Blue". These games sometimes designate files on player systems, of which must be defended, or stolen. Game servers periodically check for those flags to be intact. "Intact" could mean both existing, and unchanged. There also exist situations where players must learn the behavior of compiled programs. With conventional capture the flag competitions, there exists the following competitive objectives:

\begin{enumerate}
	\item Competitors are purposed with infiltrating existing servers
	\item Find a bug in a designated system
	\item Answer cyber security related questions, serving as education
	\item Reverse engineering binary programs, sometimes to obtain a flag
\end{enumerate}

The difference between capture-the-flag, and this system lies in how the idea of "Winning" influences the game, and different "Capabilities" of game pieces available to a player. Today, participants are rewarded points for exploiting a system, and deducted points for being exploited. Here, there is no notion of gaining points by exploitation. Participants must take it beyond simply obtaining a shell. 

During conventional CTF matches, there is a game server that periodically pings each player's system to determine whether or not a predetermined file is still on the system. If that file is not on a player's system, one assumes that the opposing player has removed the file by penetrating the player's system, and removing it. Conventional CTFs also consist of game servers that ping player resources to decide if specified services are still available. They also enable players to download binaries, and reverse engineer them for an answer to a question, of which represents a "flag".  These questions are usually related to cyber security. This is the current landscape of cyber security CTF's.

\begin{figure}[ht]
\centering
	\includegraphics[width=.45\textwidth]{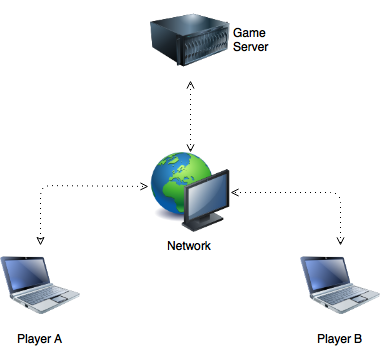}
	\caption{Capture-the-flag game design, illustrating the exchange of validations for predefined flags from players to a centralized source of truth -- the game server.}
	\label{CTF}
\end{figure}

This proposed approach is different from these conventional paradigms in that it is a methodology that simply enables participants to battle, head-to-head,  in a darwinian, survival-of-the-fittest competition. The currently existing paradigm is not a problem to be solved, and this paper aims to restate that; we merely propose a different perspective to cyber security competition. The Game Server is not focused on providing tasks, of which users must accomplish, or solve. The Game Server is instead focused on keeping track of the state of the game during battle.

\section{Proposal}

This paper introduces a system that, using a myriad of technological methods, enables competitors to directly battle one another by attempting to "destroy" one another's servers/resources -- while simultaneously oriented towards not having their own resources "destroyed". Destruction either takes on the form of rendering one of an individual player's server "unreachable", or by infiltrating an opponent's server, and grabbing a "randomly generated string" from a specific directory within that Player's server. This action serves as a Proof-of-Explotation, thus providing instant destruction of the server. Depending on the game mode, rendering a player resource "unreachable" generates an arbitrary damage amount for that given time duration defined by how long the server is "unreachable"; this is yet but one game version proposed. For this proposal, we refer to one of an individual Player's servers as a Unit belonging to a Player.

\section{Architecture}
\ii

\(p\) represents an iterator over \(P\), the set of all Players. \(i\) represents an iterator over the variable. For example \(U_{pi}\) represents, within the set \(U\), the Player (\(p\)) owning the Unit, and which Unit (\(i\)) within all of the Player's Units, \(U\)

\begin{eqnarray} \label{eq:3}
U_{pi} = {\text{Set of All Player Units}}\\
G = {\text{Set of A Game Arenas}}\\
P = {\text{Set of Players}}\\
A = {\text{Set of all Actions}}\\
C = {\text{Set of Unit Classes}}\\
R = {\text{Set of all Realms}}\\
\end{eqnarray}

A \textbf{Realm} is defined as a virtual environment in which Units of a Game exist. Let

\begin{equation} \label{eq:basic_not}
	g \in G \wedge r \in R \Rightarrow \mid G \mid \in \mathbb{R} [0,\infty)
\end{equation}

This proposal accommodates Games for only two Players, but implies \(n\) Players is possible. Also, this implies that the number of Realms in a given Game is defined as	

\begin{equation} \label{eq:realms_num}
	\forall g_i, \mid R \mid \ \in \mathbb{R} [2,\infty)
\end{equation}

\noindent
The cardinality of $R$ represents between two Realms in a Game, up to, but not equal to infinity. This is disregarding scalability, and any implementation details.
	
\begin{eqnarray} \label{eq:actions_eq}	
\forall \ a \in A, a_j = \text{Actions for \(c_n\)}
\end{eqnarray}

Each Unit, defined by its Class (\(c\)), has its own properties and capabilities. Included in its capabilities, are the Actions that the Unit can perform. \(n\) defines iterating over all Classes, and \(j\) iterates over all Actions for the specific Unit. This translates to all Units being able to perform a set of Action.

\begin{figure*}[ht]
\centering
	\includegraphics[width=\textwidth]{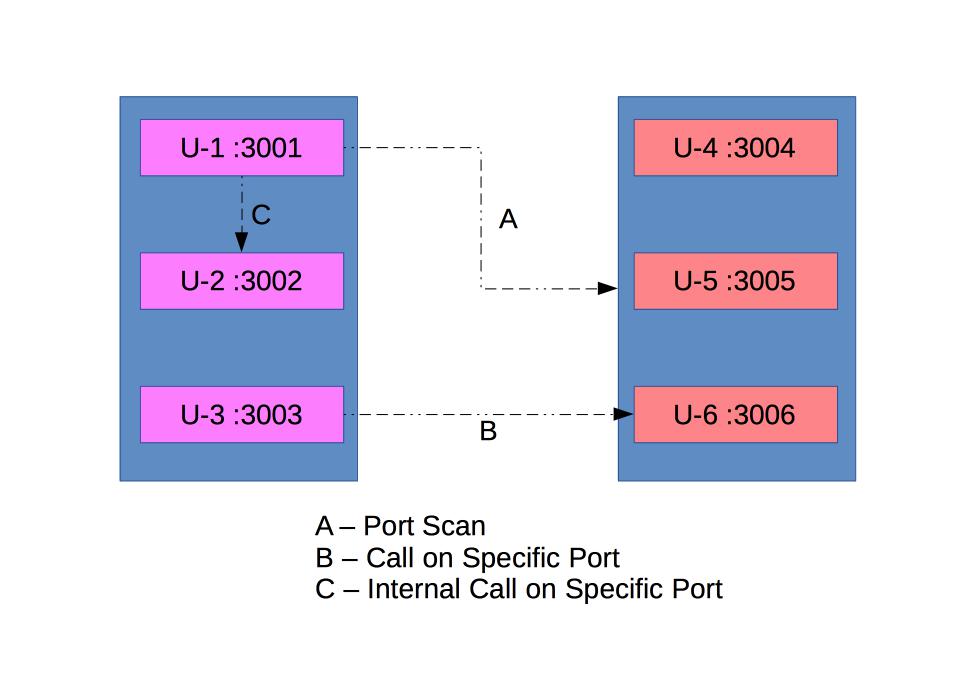}
	\caption{Unit communication potential within a Realm. Here we show two Realms, each holding three Units}
	\label{Architecture}
\end{figure*}

\section{Philosophy}

\subsection{Motivation}
\ii
The perspective of this research is motivated by multiple visions. Part of the vision stems from the subjective notion of the need to increase cyber security education in a world of ever increasing vulnerable devices being created. After a survey of the Boy's \& Girls club of America's curriculum, it is clear that DIY S.T.E.M (Science, Technology, Engineering, and Mathematics) classes are offered to participating children. However, there is a lack of cyber security education in these programs. This project is co-motivated by the gamification of cyber security in the service of attracting new practitioners to the field; agnostic of age, and background.

Another motivation for this work is to be an attempt at legitimizing the cyber security practice of "Red vs Blue" as a widely accepted Sport -- irregardless of whether it is spectator based, or private competition. In the same way that the National Basketball Association, and the Major League Baseball organization offer paths to professional athleticism, this project believes in the same potential for cyber security research.

\subsection{Penetration Testing}
\ii
Ethical hacking is not only utilized by large companies to ensure the integrity of their systems, but to also enable programmers to use their craft with the purpose of breaking into a system. As a penetration tester, your role consists of both offensive, and defensive work. Given certain situations, it is even the case that both Red and Blue actors may have to perform offensive, and Defensive tasks. The details of penetration testing help enable the existence of this proposal.

\section{Game Components}

In this section, we define the components that make up the proposal

\begin{figure*}[ht]
\centering
	\includegraphics[width=0.9\textwidth]{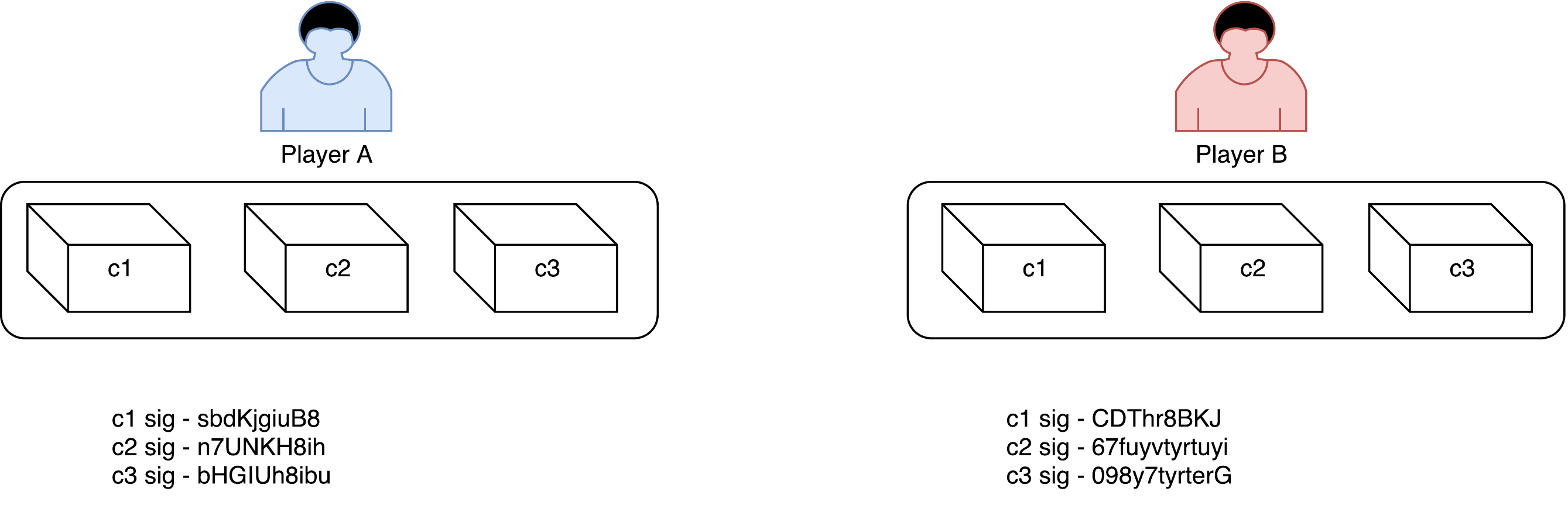}
	\caption{Each player unit has a unique fingerprint, of which regurgitation by a Player other than the owner immediately destroys the enemy Unit as a Proof-of-Exploitation}
	\label{decentralized}
\end{figure*}

\begin{itemize}
	\item Units (\(U\))
	\item Game Arena (\(G\))
	\item Players (\(P\))
	\item Realm (\(R\))
	\item Action (\(A\))
	\item Class (\(C\))
\end{itemize}

\subsection{Units}
\ii
A Unit is a single computing environment; each Unit is different, and chosen by a random process. Units bring about different capabilities, and can be used in combination with others. When they are allowed to communicate to each other, assuming they belong to the same Player, Units enable creative configurations. Units can be chosen by strategy, depending on the Game mode. The default Game mode enables random selection of Units. All Units for a given Player reside within a single Realm. Units within a single Realm can be uni-kernel, or iso-kernel. This is depending on what technology is used to implement the Units. This paper covers the comparison of containerization, and virtual machine management.

In the trials, Units are deployed with specified ports in use. These ports can either be predetermined, or determined at runtime of a Game. Regardless on when the ports are determined, neither participating Player shall have prior knowledge of the ports used. Work is required, from the Player to identify which ports are being used. Each Unit contains a predetermined service that is indefinitely running once a Unit is born. It is this service that is used to determine the "health" of the Unit.

It is also possible to have multiple services within a Unit. If multiple services exist in a Unit, the scoring mechanism must account for this difference in scoring methodology.  The trials that we perform are constrained to having only one service running at birth.

\subsubsection{Randomized Birth}
\ii
Depending on the Game mode, every Unit is selected at random from a predetermined pool of authorized Units. Each Unit is unique in nature in regards its properties. This is reinforcing the notion of unique Unit classes, and properties are discussed in detail in the section covering Gameplay.

\subsubsection{Health Paradigm}
\ii
Upon Unit birth, each unit is Born with a predetermined amount of Health; default health is 100 in the trials of this paper. Health is determined by the ability of each Player to report available services residing on their Units. To report a live Unit, the Unit is sent a network ping from its respective Player server; this network call is local to the Player server if the Units are both uni-kernel, or iso-kernel, as with the trials shown. If that ping is successful, that Unit is reported to be "up", with a status code of 200. If that ping is not successful, then the Player server reports a status code of 400, denoting that the Unit is "down". On a predefined interval, the Player host reports to the Game Server, opposed to the Game Server pinging each Unit individually. The Gameserver relies on this status for it's health scoring methods.

Determining the health of a Unit can be done in many ways:

\begin{equation}
UD = \text{Units Down}
\end{equation}

\noindent
\(UD\) is the set holding all Units that are "down" at a specific time step in the Game. 

\begin{equation}
ch = \text{Current Health}
\end{equation}

\noindent
\(ch\) is the current health of a Unit.

\begin{equation}
dc = \text{Damage Constant}
\end{equation}

\noindent
\(dc\) is the constant damage that is applied to any Unit that is an element of \(UD\), at any given time step.

\begin{equation}
s = \text{Seconds passed}
\end{equation}

\noindent
Such that a Unit's health is defined as,

\begin{equation}\label{healthexplaination}
\forall u \in UD, u_{health} = u_{ch} - s \cdot dc
\end{equation}

Depending on whether a objective-based, or time-based Game style, when the set of Units belonging to a Player all reach a \(u_{health} \le 0\), a Player is defined as \textbf{Defeated}. This is assuming a health related Game mechanism (objective-based) is being used, and that victory is determined by "last hacker alive" rules.

If a default health of 1 is used during a Game, that Game can be viewed as "sudden death" whereby Units are destroyed upon any service interruption -- even accidental interruption.

\subsubsection{Paradigm Extension}
\ii
The permutation of properties given to the Units in this proposal, as well as the permutation of other Game mechanics, is beyond an exhaustive list. Due to multiple types of Game mechanics that can be integrated with this environment, there exists many configurations that are possible. This is left to subsequent work, and the subjective vision of any other independent implementations of this initial proposal.

\subsection{Game Arena}
\ii
A Game arena is defined as the inclusion of both Player's Realms. Players have access to these Realms, and can also work externally. If the amount of Players in a Game is greater than two, \(j\), than a Game arena can encompass \(j\) Players.

\begin{equation}\label{unitsrealm}
U_G = U_R \cup U_{R+1} 
\end{equation}

The Units in any given Game is defined as the union of all Player Realms.

\subsection{Players}
\ii
A Player is defined by being a single participant in a given Game arena, owning their own Realm, and dueling against another single Player, or more. Throughout a Game, each Player has access to these "functions" -- falling under three categories (Recon, Defensive, and Offensive):

\begin{enumerate}
	\item \textbf{Recon}
	\begin{enumerate}
		\item GetOpponentIP
			\begin{enumerate}
				\item Fetches the opponent's IP address 
			\end{enumerate}
		\item GetFlags
			\begin{enumerate}
				\item Fetches a Player's own flags from their Units
			\end{enumerate}
	\end{enumerate}
	\item \textbf{Defensive}
	\begin{enumerate}
		\item Enter Unit
			\begin{enumerate}
				\item SSH a Player directly into a specific Unit
			\end{enumerate}
	\end{enumerate}
	\item \textbf{Offensive}
	\begin{enumerate}
		\item Capture Unit
			\begin{enumerate}
				\item Submit an opponent's retrieved hash string as proof-of-exploitation
			\end{enumerate}
		\item Attack Unit
			\begin{enumerate}
				\item (experimental) Enable beginners to select generic attack types, and launch them at an opponent
			\end{enumerate}
	\end{enumerate}
\end{enumerate}

\subsection{Realm}
\ii
A Realm is defined as a space containing multiple Game pieces (Units) belonging to a single Player.  Realms contain all Player data from a specific Game. In the trials, a Realm is defined as a single server belonging to a Player. All Units in the trials exist on this single server per each Player. If required, all Player Units can exist on separate servers, but this article does not cover such an approach.

\subsection{Action}
\ii
An action is defined as being a behavior undertaken by a specific Unit owned by an individual Player. Depending on player efficiency ratings, Actions can be manually generated, and programmatically constructed; they can also be fired from pre-made attack scripts for beginners. The trials focus on no pre-made attack scripts, as Players can write scripts during gameplay.

\subsection{Class}
\ii
A Class is a specific categorical label for a given Unit. The Class of a Unit determines the capabilities of that Unit, and also the available Actions that can be performed by that Unit, provided that the Unit has not been augmented by the owning Player. For the sake of these trials, the Classes for Player Units have been predefined. We also explore the concept of crowdsourcing Unit classes. A crowdsourced approach would enable for the most variety of content within a given Player Unit.

\section{Objectives}
The Game mechanics described here can take the form of many types; this paper mainly highlights three, but leads further research into others. In these three forms, Players can compete in objective-based combat, speed-based, or time-based. We enable multiple modes for the purpose of not creating an environment that only models real world scenarios, but can remain as adaptive as real-world cyber security requires. This lack of constraint is also captured in the idea of Dr. James Naismith inventing the game of Basketball. With his initial framework, it enables the opportunity to dunk a basketball today, regardless of not originally being incorporated into the game mechanics of the Sport.

\subsection{Objective-Based}
\ii
With an objective-based Game mechanic, Players can duel until all of a Player's Units are destroyed, or an arbitrary amount/percentage of a Player's Units have been destroyed. With this being a default example of an objective-based Game mechanic, it is clear that a myriad of permutations can be utilized in the service of expanding, or changing the base Game mechanics. This paper does not suggest one specific game mechanic, but rather introduces many.

\subsection{Time-Based}
\ii
With a time-based Game mechanic, Players can duel indefinitely, until a game-server-owned timer is exhausted. To determine a winner, at time exhaustion, the Game Server can take into account the sum of healths for each unit for a given player. It is also possible to take into account the amount of Units left after battle. This paper does not suggest one specific scoring mechanic for time-based matches, but rather illustrates multiple options.

\subsection{Speed-Based}
\ii
Similar to speed-based Chess, Players can begin with a predefined amount of time issued to each Player clock. By taking turns, Player time will expend per each time step taken for them to submit a move. Submission of a move can vary, as submission can be as simple as pressing "enter" within a terminal, or physically pressing a speed clock timer, as in actual speed-based Chess. In speed-based, a Player can also be defeated upon exhaustion of their respective times. This introduces a notion of acting efficiently, and quickly during gameplay.

\section{Gameplay}
The Game mechanics described here can take the form of three types. Players can compete in an objective-based environment, time-based, or speed-based. One of the purposes of this system is to enable a realistic, non-hindering, environment of which is as closely emulative of real-world cyber warfare scenarios. To support this goal we model the experience around the stages of a cyber-attack, according to the EC-Council. The stages are the following

\begin{enumerate}
	\item Recon
	\item Scanning
	\item Gaining Access
	\item Maintaining Access
	\item Covering Tracks
\end{enumerate}

Players must follow these steps in, at minimum, a haphazard manner. This means that Players must engage in all of these steps, multiple times during the course of one Game. The only surface-level exception can be "Covering Tracks"; this being because Players can be found upon interrupting an opponent's Unit. We also document a post-match judging activity, to enable a humanistic, intuitive, and artistic property to the determination of a winning party. "Covering tracks" can be considered during judging rounds; be it by manual, or automatic processes. It is documented that Games can operate with, or without these factors. One of the main purposes of this paper is to introduce the philosophy behind this ecosystem, not to merely convey dogmatic implementation details.

\begin{figure}[ht]
	\centering
	\includegraphics[width=0.5\textwidth]{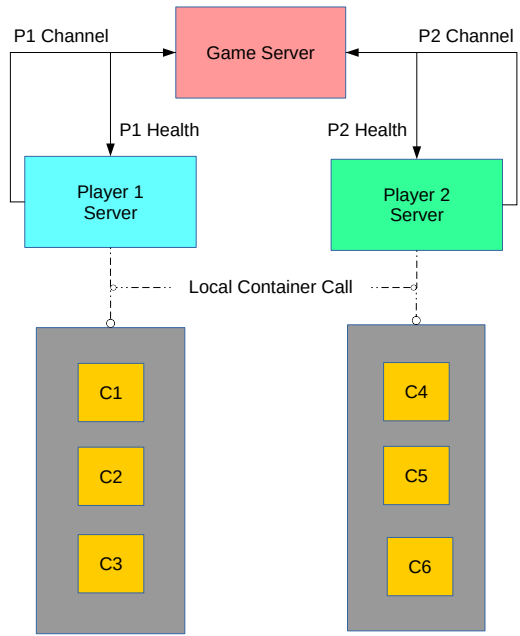}
	\caption{This is the Game mechanics from a data flow perspective, per each Game}
	\label{gamemechanicprocessmain}
\end{figure}

\subsection{Environment setup}
\ii
This paper shows multiple development versions of this Game. With one configuration, AWS, and associated products are used. With other implementations, all Game mechanics are originally developed. One configuration uses a subscription \& publishing architecture for Game Server to Player Server communication \& syncing.

\subsection{Trial Units}
\ii Without taking into account the notion of randomized Units, the following are the Units used in the Trials:

\begin{enumerate}
	\item  \textbf{Unit 1}
	\begin{enumerate}
		\item \textbf{Service} - Node Application server
		\item \textbf{Vulnerability} - system call from direct program parameters
		\item \textbf{Port} - 3001
		\item \textbf{Exploit} - remote code execution by query string
	\end{enumerate}
	\item  \textbf{Unit 2}
	\begin{enumerate}
		\item \textbf{Service} - OWASP Web Goat
		\item \textbf{Vulnerability} - Form POST is not filtered
		\item \textbf{Port} - 3002
		\item \textbf{Exploit} - Command injection
	\end{enumerate}
	\item  \textbf{Unit 3}
	\begin{enumerate}	
		\item \textbf{Service} - Damn Vulnerable Web Application (DVWA)
		\item \textbf{Vulnerability} - IP Ping service field is not filtered properly
		\item \textbf{Port} - 3003
		\item \textbf{Vulnerability} - Command injection
	\end{enumerate}
\end{enumerate}

These Units are completely arbitrary, initially used, and only used as an example of full environments with differing levels of exploitability, vulnerability, and overall capabilities. The exact permutations, and configurations for each Unit used is up to the decision of each implementation of this work. This article merely suggests a variation among many.

\begin{figure}[!htb]
	\centering
	\includegraphics[width=0.20\textwidth]{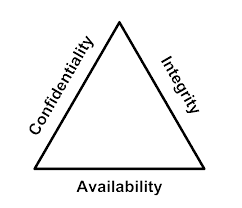}
	\caption{The CIA Triangle}
	\label{ciatri}
\end{figure}

The trial Units used all enable command execution on a target. In the real-world, not every vulnerability enables this type of exploit. With SQL Injection, an attacker can manipulate a target Database, but this does not imply shell access. To construct a system around every attack \& vulnerability type would require a change in scoring mechanism. For example, a Player Server will have to be able to determine whether a Player's database is exploited, and to what degree. Furthermore, if the vulnerability enables a Cross Site Scripting (XSS) attack, this also does not imply server access. A Game would be required to monitor the degree to which a Player's front end is being exploited. This paper does not propose a solution to monitor every attack type known. Having shell access to a server can imply complete control; this is why we focus on vulnerabilities that allow for direct command execution by some means. Referring to the CIA \cite{CIA}, the Trials focus on availability of a service, opposed to integrity, or confidentiality; this is illustrated in \ref{ciatri}.

\subsection{Crowd Sourcing}
\ii
We introduce the usefulness of crowd sourcing the internal environments for each Game Unit. This would benefit the ecosystem by introducing additional layers of entropy to the Unit generation process. While it is entirely possible for one individual to create every Game Unit environment, a crowd sourced methodology would yield Unit variations faster, and with more variability.

For the trials we model Units after industry-used platforms. We use Damn Vulnerable Web Application (DVWA), and OWASP Broken Web Application (BWA) to enable vulnerable Units. With a crowd sourced approach, the variety in the Units used increases.

\section{Demonstration}

The strategy to legitimize cyber-warfare as a Sport is directed by the production of physical, Live-events. The proposed event structure is the same as a typical, public, spectator Sporting match. The only required augmentation revolves around ensuring that Players cannot watch opponent screens, but the audience is indefinitely able to monitor both Player screens -- and a single Game Server scoring screen. To accomplish this, Player screens are tilted \( 45\degree \ge \theta \ge 0\degree \) from the orientation of a Player's opponent's \textbf{facing direction}, away from the spectating audience, while said Player is sitting at their playing station. A Player's screen is shown in the projector placed opposite of their playing station.

Each Player's shown projected screen is placed on the side of their opponent. It is possible for a Player to see their own projected screen. However, this geometric setup ensures that a Player cannot physically see their opponent's screen, in a two Player competition; the audience is still able to see both Player's screens, and the scoreboard projector behind them from the orientation of the spectating audience.

\begin{figure}[ht]
	\centering
	\includegraphics[width=0.5\textwidth]{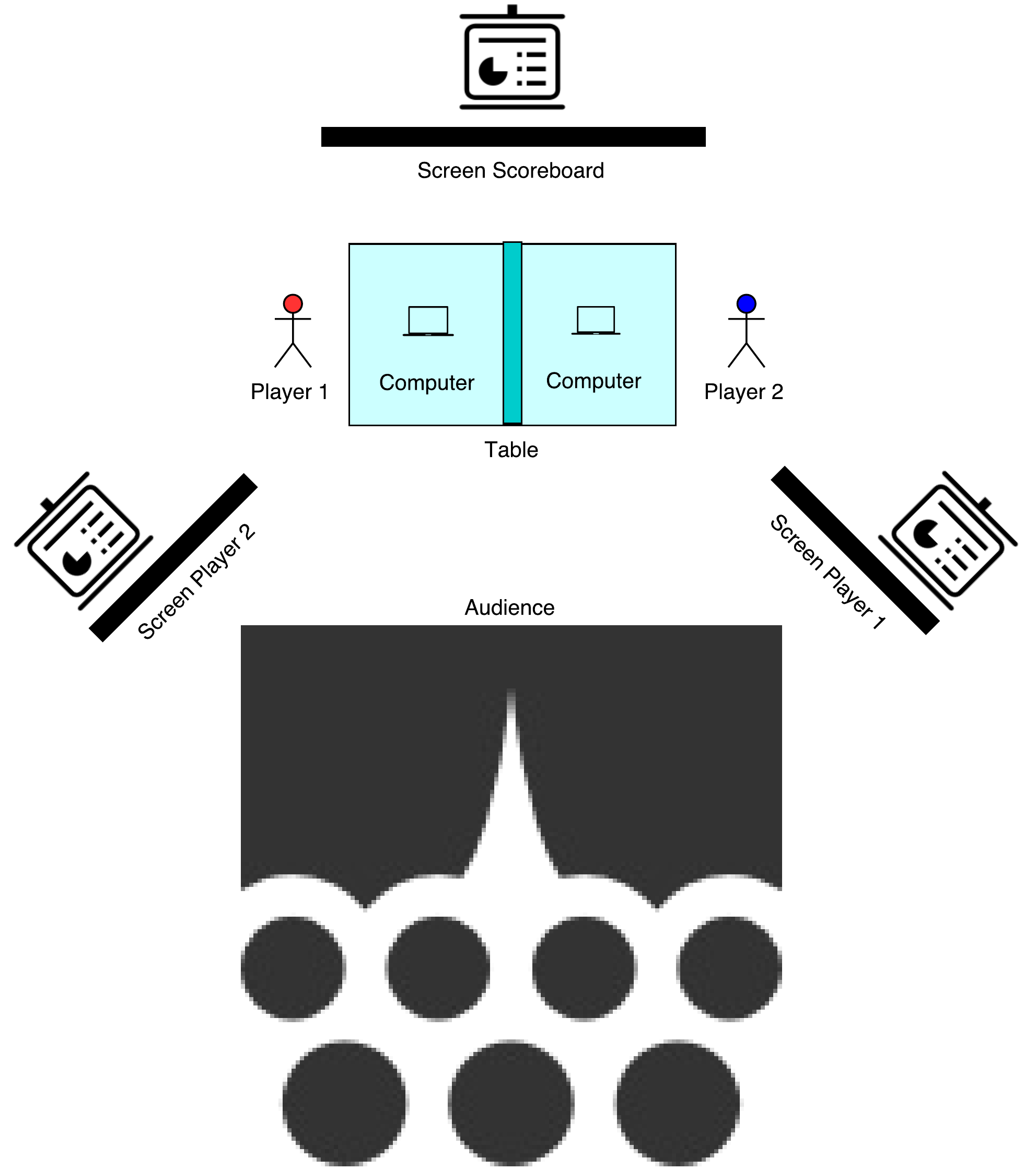}
	\caption{Top-down perspective of event structure }
	\label{eventstructure}
\end{figure}

\begin{figure*}[ht]
	\centering
	\includegraphics[width=\textwidth]{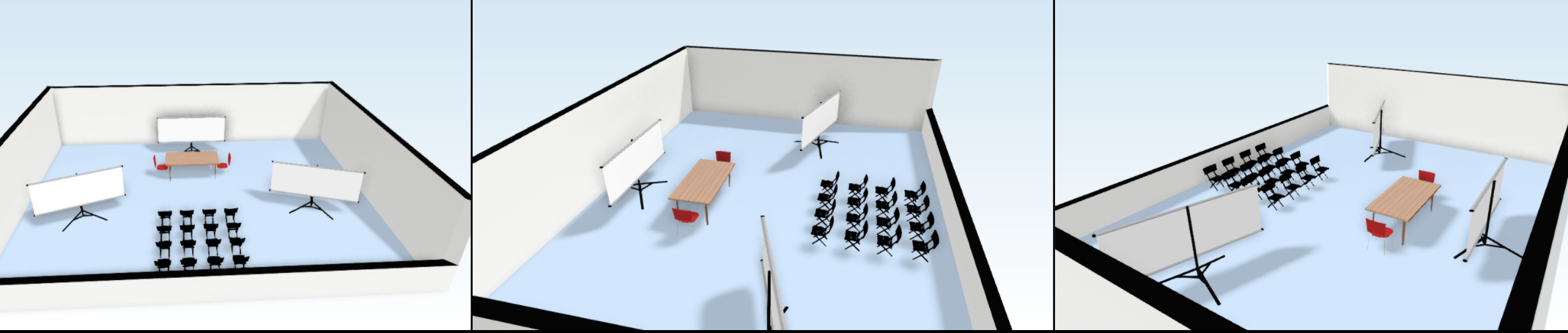}
	\caption{Event Structure from multiple perspectives}
	\label{eventstructure}
\end{figure*}

 To enable a live demonstration of the trials, the following steps must occur:
\begin{enumerate}
		\item Start Game Server on capable server
		\item Start Player servers on capable servers
		\item Ensure Players are targeting each other on the Game Server 
		\item Ensure that all Player Units are at a chosen default health
		\item Ensure Players are targeting Game Server
		\item Ensure Game Server is targeting itself for websocket UI
\end{enumerate}

\subsection{Game Server Requirements}
\ii
Assuming a centralized Game environment, the Game Server hosts a service of which is treated as a single point of truth for a Game. The Game Server in the trials utilizes the Message Queue Telemetry Transport (MQTT) protocol for the synchronization of Players and the Game Server. This choice is made because of the light-weight properties offered by the publish and subscription mechanism built within MQTT.

\subsection{Player Server Requirement}
\ii
Each participating Player hosts a Realm holding multiple Units. Units can be remote to a Player, but we focus on locality in these trials. The Player must remain updated about their own statistics, from the source of truth (Game Server). The player can also be aware of the truth about their opponent, but is not required. At a minimum, this is the purpose of a universal Game Scoreboard. If using MQTT, and if the Player Server exists on a consumer device controlled by the Player (i.e a Macbook Pro), the timing of the throughput speed for the Game-Server to Player-Server communications cannot efficiently be "real-time". 

This means that there is a tradeoff between how "real-time" the Game synchronization can be, and the speed of the Game-to-Player communications on consumer devices. To navigate this, the system is placed on three \textbf{Ubuntu 2XLarge AWS} servers. It is observed that executing the Player environments on high-performance servers yields actual real-time experience, but removes control of the Player environment from the local possession of the actual Player.  Additional implementations can use controlled REST methods for Game synchronization, but a "real-time" experience is preferred in these trials.

\subsection{Player Targeting}
\ii
The Game Server must also be aware of identity of each opponent, relative to the Player. In other words, the Game Server has to be told which Player is the opponent of another Player. In these trials, each Player is paired with a single opponent. In three-way competition, a Player is assigned two opponents, and so on. Enabled by Player targeting, a single Game Server can mediate multiple, independent Games. The scenario of duplicate Player names is not examined, but is discouraged for the reservation of Player identity "uniqueness".

\subsection{Health Default}
\ii
In a simple demonstration, we begin a Game with each Unit starting at 100 health points. A calculation for the \(health\) of a Unit is shown in equation \ref{healthexplaination}, and is dependent on how many times a server has been flagged as locally-unreachable by a Player service check. 

This is a convenient Unit health with which to begin. A choice in the trails is also made to begin certain Games with a default Unit health of 1. This enables a "sudden death" Game type, whereby Units are destroyed upon any service interruption. 

\subsection{Game Server Subscription}
\ii
Each Player must publish, and subscribe to the Game Server holding their source of truth, unless in a decentralized Game version. While being subscribed to the same Game Server, the Game state is shared between all parties involved; each Player, and a Game Server. A Game is not dependent on a publish, and subscribe structure, but this structure is conducive to the trials demonstrated in this work.

\section{Experiment Design}

\subsection{Questions}
The following is an exhaustive list of our experimental questions:

\begin{enumerate}
	\item Is it possible to develop a system \& methodology that enables head-to-head competition based on cyber security?
	\item What performance issues arise when enabling such a system with virtual environments?
	\item What features of the real-time environment are of interest to warrant analytics?
	\item What is the likelihood of a modern cyber-security professional desiring to participate in such a competition?
	\item What does a comparison in terms of energy consumption look like with other Sports?
	\item How does skill, and experience play a factor in such an environment?
	\item What are the challenges in this area for Artificial Intelligence work?
	\item How does the environment change if the source of truth is decentralized?
\end{enumerate}

\subsection{Hypotheses}
The following is an exhaustive list of experimental hypotheses regarding the questions mentioned heretofore:

\begin{enumerate}
	\item The energy consumption of such an activity takes the same amount of energy as at least one given sport, during a span of 1 hour.
	\item This Game environment creates a competitive environment that induces measurable stress on any participant.
	\item A player with no cyber security skill set cannot readily play this proposed game at a generally competitive level.
	\item Training an Artificial Agent to play this game is not intractable.
	\item Incrementally playing this proposed game is turing decidable for an Agent.
	\item Spectators with no knowledge of cyber security work will find this activity appealing to observe.
	\item A decentralized, and centralized Game structure yield the same competitive environment.
\end{enumerate}

\subsection{Trial Game Ecosystem}
\ii
For our experiments, we enable three server environments. Two for the Players of the Game, and one for the scoring Game server. The Player servers are subscribed to a channel providing data on their own health, and another channel providing data on their opponent's health. The Game Server acts as a traffic cop for the Game data. In a centralized structure, this ensures that all real-time calculations are from one source of truth -- the Game Server.

\subsection{Tools}
\ii
We develop an MQTT client on Player Realm servers to enable real-time, uninhibited Game scoring. We also develop executables that enable a Player to submit commands to the Game Server. For example, we expose an executable in C++ that when invoked with a string as a parameter, contacts the Game Server, and attempts to "submit" an opponent flag, proving Player exploitation to the Game Server.

AWS is used in a myriad of ways during this project. The initial experiment trial, covered in the section on Game Versions \ref{gameV} uses AWS in a black-box manner. Due to the technology requirement being so heavy, that approach is not used any further. AWS is used in these trials for access to computationally efficient servers within which to host both Player servers, and Game Server. This proposal is in no way dependent on AWS, or any service that is similar.

\section{Experiments}
We conduct various experiments highlighting variables of interest to the research team. Among these interests are the \textbf{speed} with which the Players operate, the psychological \textbf{stress} levels observed on the Players during gameplay, the \textbf{consistency} that Players exhibit when typing \& working on their system, the \textbf{error rate} observed from Players during gameplay, \textbf{characters-per-minute} (ChPM), \textbf{commands-per-minute} (CoPM), and \textbf{commands-per-entry} (CoPE). These properties of the Game are simply the properties chosen for the trials of this work. Future implementations should experiment with additional properties. After a Game is finished, a Player's style-of-play can be profiled using these metrics.

\subsection{Speed}
\ii
While actively working during a Game, the speed of typing, and command entry is measured in real-time. This is similar to calculating words-per-minute from the perspective of generic computer typing tests. Speed can be indicative of how comfortable a Player feels with the environment used, and how fast they can type during the performance of work.

\subsection{Stress}
\ii
The focus around stress involves the question of "What forms, and levels of stress does an average participant endure during an ethical hacking exercise?". It is a focus of this paper to support cyber warfare as a legitimate sporting activity. We observe different indications of stress, and behavior associated with it. If one could closely compare existing sporting activities with the activities of ethical hacking, it can serve as the initiation of a conversation on the matter -- at the very least.

By summarizing the results of studies that measured the circulating levels of stress hormones before and after individuals were exposed to various situations that were deemed to be stressful (e.g., air-traffic controllers or parachute jumping), Mason (1968) was able to describe three main psychological determinants that would induce a stress response in any individual exposed to them. 

Using this methodology, he showed that in order for a situation to induce a stress response, it has to be interpreted as being novel, and/or unpredictable, and/or the individual must have the feeling that he/she does not have control over the situation. Recently, another determinant was added to this list -- namely a threat to the ego. Although this work led to a general debate between Selye and Mason (Selye, 1975), further studies confirmed that the determinants of the stress response are highly specific, and therefore, potentially predictable and measurable \cite{stress}.

\begin{table}[]
\centering
\caption{Participant age, identification code, gender, years professional, and testing start time for physiological experimentation}
\label{parttable}
\begin{tabular}{llllll}
Code 	& 	Age 	& 	Sex 	& 	Years 	& 	Testing Start Time	 	&  \\
P01  	& 	24  		& 	m   		& 	7     	&    9:30am     				&  \\
P02  	& 	26  		& 	m   		& 	13     	&    1:30pm               		&  \\
P03  	& 	21  		& 	m   		& 	1     	&    12:30pm  	 	            & 
\end{tabular}
\end{table}

We measure Cortisol levels, heart rate, and breathing frequency during gameplay. We also monitor EEG channels. We combine all stress measurements, except EEG data, synced by time, onto a single visualization in figure. Table \ref{parttable} shows the three participants included in the study.

\subsubsection{Measurement Equipment}
\ii
We procure multiple pieces of equipment for physiological measurement. For Heart Rate, we purchase a Fitbit Charge ll fitness band. For Cortisol monitoring, we purchase multiple ZRT Adrenal Stress Profile Hormone Imbalance Home Test Kits by Genova Diagnostics \cite{cortprof}. This is a salivary cortisol test that is taken by Players at multiple times during the day. For Breathing frequency, we procure an RMN-204 Respiration Monitor \cite{breath}. This enables for the measurement of depth, and frequency of Player breathing. Lastly, for EEG measurements, we rent the Low-profile MicroCel Geodesic Sensor Net from Electrical Geodesics Incorporated (EGI) \cite{egi} with the necessary software from EGI.

\subsubsection{Saliva}
\ii
Many assay techniques are available to quantify free cortisol from saliva samples. The most common assays are radioimmunoassay (RIA), time-resolved immunoassay with fluorometric detection (DELFIA) and enzyme immunoassay (EIA). These techniques rely on the principle of competitive binding between free cortisol and reagents. Correlations betweens concentrations yielded from these techniques depend on the type of population tested (clinical vs. healthy) and on the range in concentrations assayed (Addison vs. Cushing) (Raff, Homar, \& Burns, 2002; Raff, Homar, \& Skoner, 2003) \cite{stress}. 

Therefore, one should use some caution and consider the type of assay used when comparing values obtained from one study to another study. The choice of one technique over another depends not only on the prices of the chemical kits, and availability at laboratories, but also on the percentage of inter and intra-assay coefficient of variations. Briefly, inter-assay variations refer to the variability related to the assay between runs, while intra-assay variations refer to the variability within runs. Further information regarding assay techniques can be obtained from Dr. Claire-Dominique Walker and Dr. Michael Meaney \cite{stress}. With our approach, a 7 day time buffer is required between collecting a sample, and receiving the results of the level of Cortisol concentrated within it. The results are received from the supplier of the Salivary test kits. Upon collection, the time of collection is carefully recorded for each sample. This is crucial for the time-series reconstruction of the Cortisol levels. The lead time is significant due to the lack of salivary, and Lateral Flow Assay (LFA) expertise.

\begin{figure}[ht]
\centering
	\includegraphics[width=0.5\textwidth]{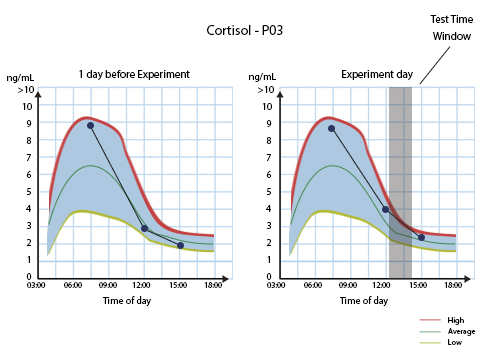}
	\caption{Cortisol measurements for participant P03, on the day of, and the day before experimentation. Time of gameplay window is highlighted on the day of experimentation. }
	\label{cortisol}
\end{figure}

As a control, we also require participants to measure Cortisol levels for the day before testing. This is used for comparison to the experimentation day. Figure \ref{cortisol} illustrates Cortisol levels for participant P03 in the trials. It is to be noted that the participants are not engaging in these experiments over the course of a full day, but rather less than 2 hours. This hinders a direct time analysis of Cortisol levels, but nonetheless enables a glimpse into Player stress.

\subsubsection{Heart Rate}
\begin{figure}[ht]
\centering
	\includegraphics[width=0.45\textwidth]{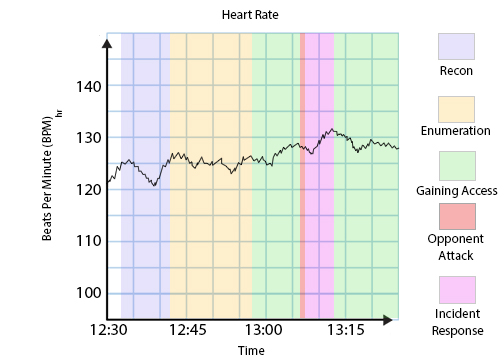}
	\caption{Participant P03's beats per minute displayed with associated hacker phases}
	\label{heartrate}
\end{figure}

\ii
Using a Fitbit Charge ll, we monitor the Beats Per Minute (BPM\(_{hr}\)) within the Fitbit mobile application for iOS. Heart Rate, unlike saliva, can be retrieved and calculated in near real-time. Figure \ref{heartrate} shows participant P03's heart rate, with associated hacker phases.

\subsubsection{Breathing Frequency}
\begin{figure}[ht]
\centering
	\includegraphics[width=0.5\textwidth]{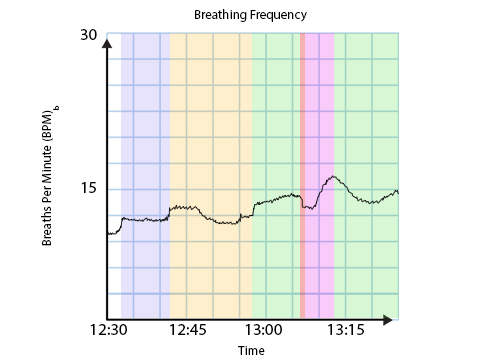}
	\caption{Shown is the breathing rate of participant P03 during Game experimentation}
	\label{breathing}
\end{figure}

\ii
Using the RMN-204 Respiration Monitor, we measure the Breaths Per Minute (BPM\(_b\)). The monitor measures the amount of breathing occurring within a specific timeframe, and calculates the projected BPM\(_b\). We observe faster breaths associated with heighten levels of stress, and anxiety. Figure \ref{breathing} illustrates the increase in breath frequency during many hacker phases. Figure \ref{breathing} also shows deep breaths being taken upon the participant realizing one of their Units have been exploited.

\subsubsection{Electroencephalography }
\begin{figure*}[ht]
	\centering
	\includegraphics[width=\textwidth]{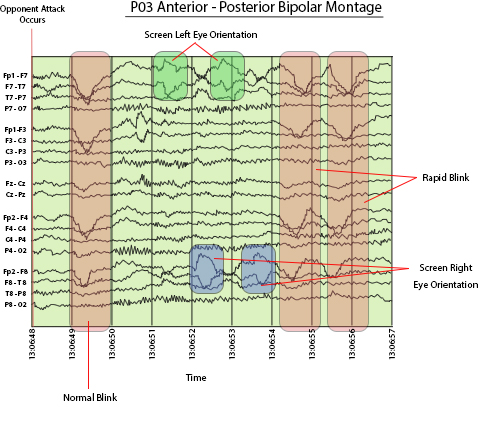}
	\caption{Anterior - Posterior Bipolar Montage of participant P03 immediately after an opponent exploitation}
	\label{eeg}
\end{figure*}

\ii
Electroencephalography is the measurement of electrical activity in different parts of the brain and the recording of such activity as a visual trace (on paper or on an oscilloscope screen). Using the Low-profile MicroCel Geodesic Sensor Net, we visualize participant P03's measurements in figure \ref{eeg}. We also highlight the hacker phases upon which the participant is focused during the readings. We use the Anterior - Posterior Bipolar Montage representation of the EEG channels. We observe the participant immediately after an opponent attack. We observe rapid eye movement between the outer edges of the screen, followed by two blinks. Previous to this time window, we observe consecutive normal blinks. It is proposed that the realization of the opponent attack serves as the stimulus causing the observed eye movement. We reference this same point in time in figures \ref{breathing}, and \ref{heartrate}. Cortisol levels are not able to be this granularly correlated due to the collection method used.

We observe increased physiological metrics following the instance of these eye movements. This paper does not exhaust methods of validating such claims. It is a purpose of this paper to introduce methods of monitoring Player biology, and is an attempt to appeal to the richness of analytics enabled by the proposed approach.

\subsection{Metric Aggregation}
\begin{figure}[ht]
	\centering
	\includegraphics[width=0.5\textwidth]{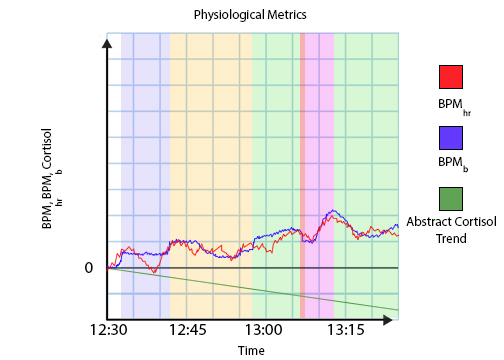}
	\caption{The superimposition of various physiological measurements through time}
	\label{agg}
\end{figure}

\ii
We aggregate breath frequency, heart rate, and cortisol measurements into one time-series plot. We sync the plot by time, and highlight which hacker phase the participant is in at the moment. For the Cortisol readings, we must use fewer data points, due to the 7 day time buffer between saliva capture, and result retrieval. However, this is handled by the time recording upon sample collection. We maintain the time synchronization between data points by superimposition, and scale merging. Figure \ref{agg} illustrates the aggregated data points from these physiological tests.

\section{Gameplay Metrics}
We formulate the metrics that are monitored during gameplay. In this section, we highlight consistency, error per minute, characters per minute, commands per minute, and commands per entry.

\subsection{Consistency}
\ii
We highlight the behavioral consistencies that Players exhibit, when available. Consistency only measures the usage of specific commands, while in certain situations. This can be a subjective measure depending on the Game version. To measure consistency, we observe the variety amongst a Player's commands. We do not attempt to formalize which is better for use during gameplay. We simply observe the type of command consistency that exists during gameplay. It should be noted that it may not necessarily be better for a Player to use different types of commands to accomplish the same thing. However, this is subject for another research project.

\subsection{Error Per Minute (EPM)}
\begin{figure}[ht]
	\centering
	\includegraphics[width=0.25\textwidth]{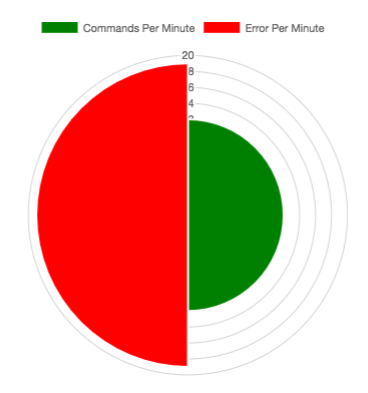}
	\caption{Error rate (red) visualized with commands per minute (green), as calculated in equation \ref{eq:cmd_time}, used to monitor Player behavior. }
	\label{commandpermin}
\end{figure}

\ii
We analyze player typing errors by highlighting frequency of "Delete", and "Backspace" appearing in a key logger. This is indicative of a player entering an erroneous command, or character. This can be done by typing error, or error is logical approach, by a Player, at a given time during gameplay. EPM is shown in figure \ref{commandpermin}.

\subsection{Characters Per Minute (CPM)}
\begin{figure}[ht]
\centering
	\includegraphics[width=0.5\textwidth]{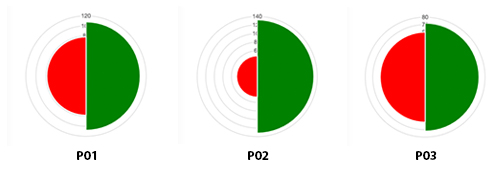}
	\caption{cumulative CoPM and EPM, at the 15-minute mark during experimentation for participants P01, P02, and P03 respective}
	\label{playercomfort}
\end{figure}

\ii
CPM is keeping track of how many characters a Player types per minute; this is important for typing speed. It is possible to monitor Player eye position while typing for the purpose of gauging "keyboard comfortability". It is observed that both participants, P01 and P02, were able to type without looking at their keyboards. P01, and P02 had a higher average amount of CoPM, and CPM as seen in figure \ref{playercomfort}. This is in comparison to P03, of whom did not possess the same ability.

\subsection{Commands Per Minute (CoPM)}
\ii
CoPM keeps track of the amount of full \& valid commands a Player inputs per minute. As a programmer given the task of attacking a system, speed is an influential metric for the effectiveness of a programmer. With this in mind, the system monitors the speed with which a Player is entering commands into their Game environment. Visually, we plot the amount CoPM, as a function of time. CoPM is also shown in figure \ref{eq:totalamountofcommands}. We represent the set of total commands, at a given time \(t\), as \(C_t\),

\begin{equation} 
\label{eq:totalamountofcommands}
	\{C_{t} \mid \text{Set of commands at } t \}
\end{equation}

\(G_{ET}\) represents the time that has passed in the Game, at the point in time in which CoPM is being computed,

\begin{equation} 
\label{eq:elapsedtime}
	G_{ET} = \text{The elapsed time passed in Game}
\end{equation}

\(C_{PM}\) holds the CoPM results. This is mentioned for the purpose of thinking programmatically,

\begin{equation} 
\label{eq:cmd_per_min}
	C_{PM} = \text{Commands per minute}
\end{equation}

\(C_{TS}\) is the number of commands entered in the most previous, 10-second window. This is only used when calculating the CoPM in real-time, opposed to retrospectively.

\label{eq:cmd_tensecond}
	\begin{equation} 
\label{eq:totalcommandset}
	\{C_{TS} \mid \bigcup \limits_{ t=0 }^{ t-10 } C_{t} \}   
\end{equation}

We observe the cardinality of \(C_{TS}\) for the amount of total commands entered in the last 10 seconds,

\begin{equation}
\label{eq:tenseccard}
	|\ C_{TS}\ |
\end{equation}

One method of calculating \(C_{PM}\), is to divide the total amount of commands, by the amount of time elapsed during the game

\begin{equation} 
\label{eq:cmd_time}
	C_{PM} = \frac{ C_{T} } { G_{ET} }
\end{equation}

a second way to calculate CoPM would be to count the amount of valid commands per every 10 seconds, and multiply by 6, meaning the amount of 10 second intervals existing within a 1 minute time duration, or \(\frac{60}{10}\).

\begin{equation} 
\label{eq:cmd_time}
	C_{PM} = 6\ |\ C_{TS}\ |
\end{equation}

Because of the relationship between the time \(t\) in the game, and the commands, it is implied that their should be a method/function that performs a computation on \textbf{Time} \& \textbf{Command}, to calculate the \( \frac{commands}{minute} \) for a given player (\(p\)). As example, it can be as:

\begin{verbatim}
int CMDPerMin(Time t,CMD cmd, realTime = 0){
    if(realTime){
      CMD[] c = CmdsByRange(t,t-10) 
      int numberOfCommands = sizeof(c)
      return numberOfCommands * 6
    }else{
      CMD[] c = getTotalCMDs()
      int numCMD = sizeof(c)
      return numCMD / t.elapsedMins()
    }
}
\end{verbatim}

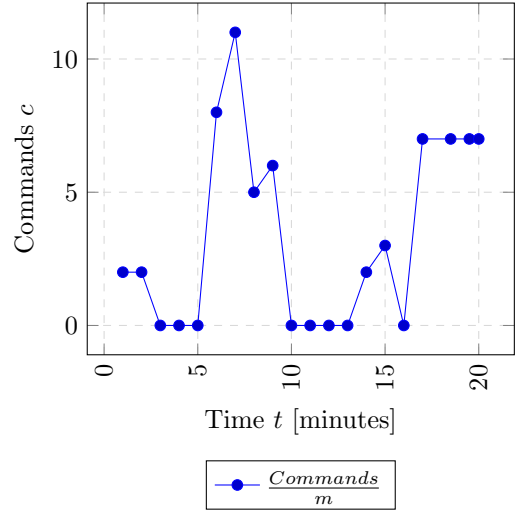
\begin{figure}[h!]
  \begin{center}
    \begin{tikzpicture}
      \begin{axis}[
          width=0.9\linewidth, %
          grid=major, %
          grid style={dashed,gray!30}, %
          xlabel=Time $t$, %
          ylabel=Commands $c$,
          x unit=minutes, %
          legend style={at={(0.5,-0.3)},anchor=north}, %
          x tick label style={rotate=90,anchor=east} %
        ]
        \addplot 
        table[x=column 1,y=column 2,col sep=comma] {table.csv}; 
        \legend{ \(\frac{Commands}{m}\) }
      \end{axis}
    \end{tikzpicture}
    \caption{The amount of commands per minute passed in the game whereby a trough indicates a player pause in typing action. This is captured from participant P03 during experimentation}
  \end{center}
\end{figure}

\subsection{Commands Per Entry (CoPE)}
\ii
This keeps track of the number of commands, on average, that are chained into single terminal command entries by a Player. CoPE enables insight into how many commands a Player is inserting as "one-liners". This appears as player-separated commands, typically by a semi-colon ";". This measurement enables a psychological view into a Player's behavior and play style.

\subsubsection{CoPE Psychology}
\ii
In terms of measurable behavior, high CoPE occurs when a player either already knows the effect that the given commands will have on the environment, or if they are attempting to string commands together. A Player that is unsure about the environment is more reluctant to string multiple commands together, unless there is a strategy as motivation. This strategy could be aligned with wanting to see the capabilities of a given environment, not caring about the insertion of incorrect, or nonexistent commands. 

A Player that is unsure about their environment, is more likely to step through their processes; this is to closely observe the relationship between their input into the environment, and the output from the environment. This paper does not substantiate these claims, and does not intend to do so at the current moment. However, this will either be substantiated, or disproven as more Games are played by various unique Players. This has only ben observed among the experiment Participants.

\section{Game Versions} \label{gameV}
Due to the exploratory nature of this work, we experiment with a myriad of Game versions. We use both black box, and proprietary solutions. This section covers the variety of approaches, and introduces both the positive, and negative factors of each decision.

\begin{figure}[ht]
	\centering
	\includegraphics[width=0.5\textwidth]{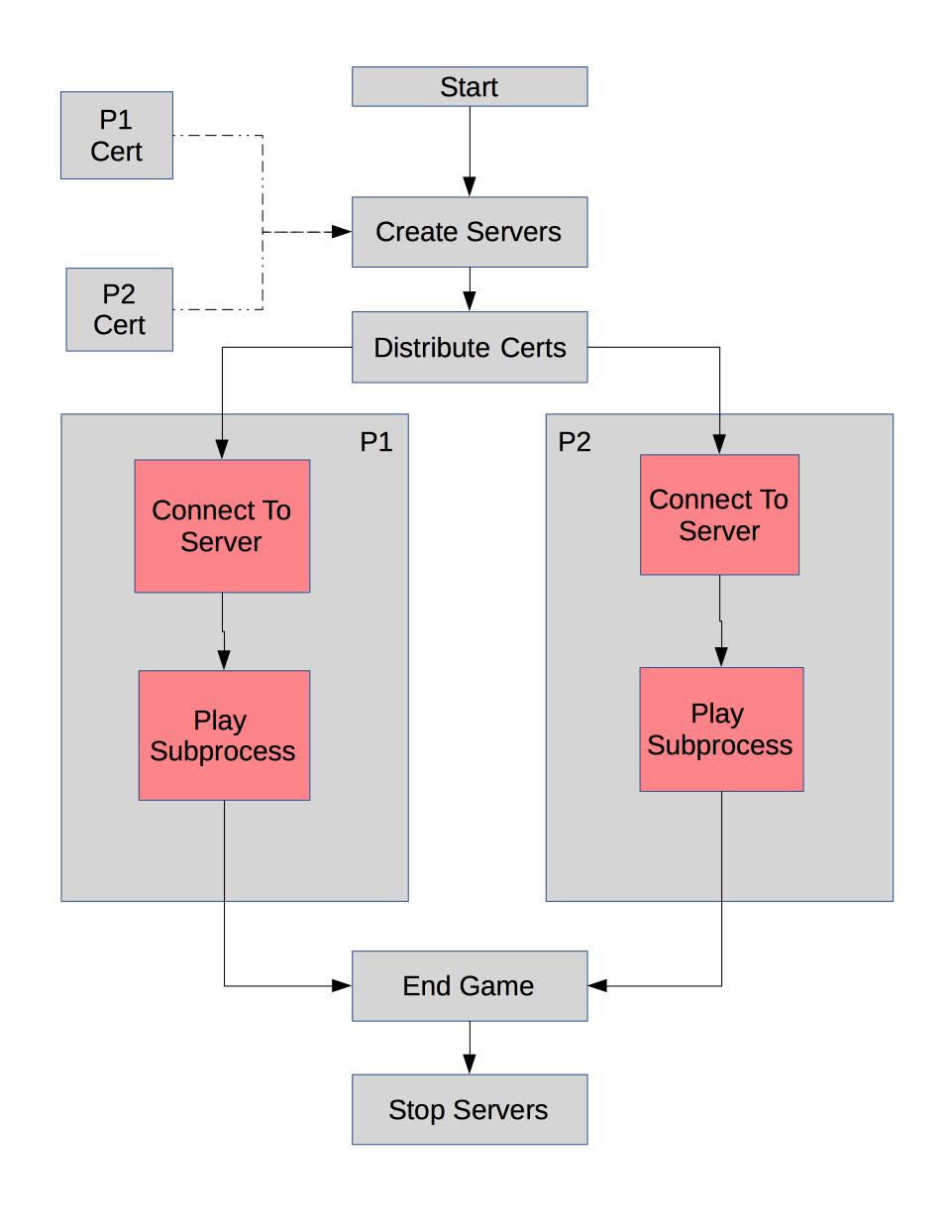}
	\caption{Early AWS Experimental Process }
	\label{game_flow}
\end{figure}

Using the Amazon AWS version comes with its own benefits, challenges, and disadvantages. AWS comes with built-in server orchestration tools. We originally use AWS, AWS-CLI, GOLang, and MongoDB, player certificates, and SSH sessions to enable Games. This configuration is conveyed in figure \ref{game_flow}. The problems with this proof of concept include risking a dropped SSH connection during gameplay. This shows a need for "local" SSH sessions, and also raises the question about the "real-world"ness of a Game, and what should be suitable. If we are aiming for real-world simulation, then SSH sessions are not rare to be dropped. For the purpose of the Game, the Game environment could only include Player A's local machine, Player B's local machine, and a Game Server. We eventually use both local, and remote Player environments.

As mentioned before, there is an issue of "hosing" a Player's computer, since the Game Server is sending, and fetching data in real-time via a specific protocol -- in this case MQTT. To circumvent this consumption of resources, we elevate performance on Player servers. We enable two remote servers with full game environment setups included (all Game dependencies installed). Upon creation of a Game, the Players are given SSH access to their servers, to which they are expected to SSH, for the purpose of adjusting \& augmenting their own Units, submitting a "Capture" command, and if they just want to use more "capable" resources. The main drawback of this, like stated before, SSH sessions can drop at the most inopportune moments during gameplay. This highlights a tradeoff between resiliency of gameplay, and ease of play. However, for the actual development of the game beyond a proof-of-concept, we choose less predefined solutions. This enables customisation of all Game mechanics.

\begin{figure*}[ht]
\centering
	\includegraphics[width=0.8\textwidth]{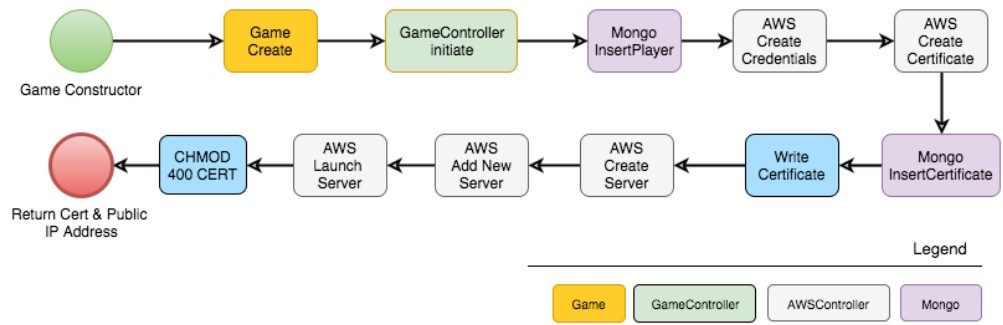}
	\caption{The game setup process for a centralized game structure experiment \#1. Process returns credentials needed to SSH into their player server \(s\). This specific proof of concept used Amazon AWS, and associated platforms.}
	\label{decentralized}
\end{figure*}

\subsection{Player Server Locality}
\ii
Another Game version involves a Game server hosting every Unit within each opponent Realm. This relinquishes full control of the environment to the Game Server. Another configuration involves the remote of every Player Unit.

\begin{enumerate}
	\item Remote
	\begin{enumerate}
		\item \textbf{Pro}: Multi-device, Multi-platform
	\end{enumerate}
	\begin{enumerate}
		\item \textbf{Con}: Player can lose connection during Game
	\end{enumerate}
	\item Local
	\begin{enumerate}
		\item \textbf{Pro}: Organizational benefit of bundle download
	\end{enumerate}
	\begin{enumerate}
		\item \textbf{Con}: Device throughput/resource limitation
	\end{enumerate}
\end{enumerate}

\section{Protocol}
These trials are implemented using MQTT, which operates on top of TCP. This leads to observations of Player workstation limitations, whereby the frequency of the packets from MQTT begin to consume more computer resources, as the Game progresses. We test this on two, consumer grade, Macbook 13'' laptops, and two, consumer grade, MacOS desktop computers. The resource consumption is consistent with at least 2 Players playing one Game. 

We also propose a method of utilizing a custom protocol during gameplay. This protocol is used to enable a smaller data footprint during gameplay, leading to lower bandwidth required for base Game mechanics. Using MQTT already uses a very small protocol scheme. However, the data is broadcasted very frequently, due to the nature of the protocol. This is however configurable, but we do not exhaust the configurations in this work. Without needing to change MQTT's polling nature, we can devise a scheme to only transfer the data deemed important for the Game to operate. Creating individual MQTT topics for each Player, and Unit combination is a tradeoff between size of protocol overhead, and ease-of-play on consumer devices. This is not thoroughly covered in this work, and assumes a capable MQTT broker server.

\section{Unit Environments}
\indent
We develop both a light-weight, and simulated real-world implementation. The light-weight version is mainly purposed for rapid education, and training. For competition environments with dedicated environments, the real-world simulation emulates live scenarios. Both versions can be rapidly deployed, and decommissioned. In this section we highlight the key difference between the proposed lightweight, and real-world approaches.

\subsection{Containers}
\ii
Deploying the system in a containerized environment, allows the system to be played in any environment supporting the containerization service. Using containerization allows for additional architecture configurations without the overhead of using virtual machines (VM). These configurations can differ from where the Game data is stored, how the Units are allocated, and where the Units reside during gameplay. Configurability is not unique to containerization, as VMs are configurable as well. 

We point out that deploying this system using containers instead of VMs characterizes this Game as a light-weight, but still dynamic simulation environment. The light-weight categorization is due to the notion of each Unit sharing a kernel with another Unit (unikernel). This assumes each of a single Player's Units exist on a single host. Containerization is conducive to educational, and training purposes without unnecessary overhead.

\subsubsection{Docker}
\ii
This project's light-weight implementation uses the Docker engine \cite{docker}. We observe limitations of the platform; most of which are not covered in this paper. One substantial issue that exists when using the Docker engine arises when the interruption of a base process within a container occurs. For example, if we create a container to fundamentally expose a Node JS server, upon the forced interruption of the Node JS process, SSH, and any further communication is hindered. This project does not investigate this problem further, but it is believed to be caused by interrupting a process of which the container is originally generated to perform. It is to be noted that these Units are created using specific Dockerfiles. It is the Dockerfile that configures the base processes to run inside the Unit. We create light-weight Units with the following structure. 

\begin{verbatim}
    docker run -d -p <port>:<port> <unit>
\end{verbatim}

The port is the port upon which the base process runs.

\subsection{Virtual Machines}
\ii
Deployment using VMs enabled for truly isolated environments. The benefit coming from a hard partition that a CPU when using a VM. The notion of not sharing a kernel (isokernel) also allows for an experience that resembles a real-world environment. The tradeoff to highlight here is that these isolated boxes require more overhead on the host being used.

\subsubsection{Vagrant}
\ii
For the real-world simulated environment in this project, we use Vagrant. This enables the research team to deploy fully isolated VMs on a given host. Using vagrant allows us to mitigate the "lost of communication" problem observed with Docker. This also enables for reliable SSH capabilities for Players.

\begin{verbatim}
config.vm.provision :shell, path:"start.sh", 
run:"always", privileged:p
\end{verbatim}

we use p, as a boolean, to denote Unit privileges. The use of Docker, and Vagrant are different, and warrants development of separate Unit creation processes. Again, this ultimately depends on the purpose of a Game. However, the separation, and sharing of kernels among Units is used as the delineation of light-weight, and simulation.

\section{Source of Truth}
The Game mechanic has a potential for various paradigms in terms of where the "source of truth" lies. You can have a centralized source of truth with a single Game Server, or you can make use of a decentralized source of truth by pure peer-to-peer connections.

\subsection{Centralized}
\ii
Here we will examine the Game with a centralized architecture. With a centralized architecture, you only require one Game Server. This Game Server is a traffic router for Game data traveling to each Player. Conventional CTF games adopt a centralized game structure. Although, conventional scoring is different from scoring in this proposal, a centralized Game structure remains the same across implementations -- as far as "source of truth" is concerned.

\subsection{Decentralized}
\ii
Here we will examine the Game with a decentralized architecture. With a high amount of concurrent Players, each scoring computation is validated by peer nodes. This enables consensus to take place within Games. Theoretically, every Player is also a Game Server, and thusly a "source of truth". The disadvantages of a centralized system is the inherent requirement to trust the integrity of that single source of truth. During a physical event, with officials present, this may not be a problem. However, once this system is being used by a high volume of Players at one time, this requirement exposes an integrity-based vulnerability within the system itself. This also warrants the exploration of different decentralized Game structures.

\subsubsection{Blockchain Validation}
\ii
Using a Blockchain for various Game information, the system can self-validate information, and also validate that a specific Player sent a specific action; similar to Bitcoin's proof-of-work. The issue at the time is the speed with which reliable proof-of-work takes. The aim is to submit Player Actions onto the ledger. This way validation happens with a form of consensus. This is to be used during Games with multiple Players. In a blockchain ledger, the blocks will fundamentally comprise of representations (hashes) of the following:

\begin{verbatim}
{ timestamp: <date> <time> <timezone>,
  ip: <ip address>,
  playerName: <player name>,
  cmds: {},
  units: 
   [ { code: 200,
       id: <unit 1 identity hash>,
       health: 100,
       port: <port 1> },
     { code: 200,
       id: <unit 2 identity hash>,
       health: 100,
       port: <port 2> },
     { code: 200,
       id: <unit 3 identity hash>,
       health: 100,
       port: <port 3> } ] 
}
\end{verbatim}

\begin{figure}[ht]
	\centering
	\includegraphics[width=0.5\textwidth]{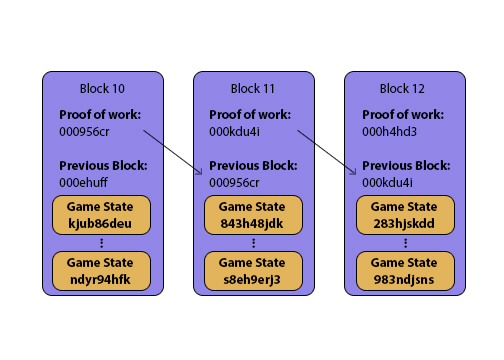}
	\caption{The storage of Game state being stored in a blockchain ledger}
	\label{blockchainstructure}
\end{figure}

Each hash, along with other information, is stored in each block. This enables a distributed history of "transactions" within a Game. A transaction can be thought of as a single record of total Game state at \(t\). We do not explore different consensus algorithms for blockchain usage during peer-to-peer gameplay, but significant work is warranted in this space. Figure \ref{blockchainstructure} shows the Game state being stored within a blockchain ledger. Game state includes all properties needed for a Game, and Player commands entered throughout the Game.

\subsection{Advantages}
\ii
\indent The benefit of a decentralized nature is the support for a gaming environment that is governed by the system itself, not a central authority. Digital history has flourished with the rise of peer-to-peer systems. The proposed system could inherent the same traits with a decentralized structure.

\subsection{Disadvantages}
\ii
The problems with a decentralized structure begin to arise regarding the environment dependencies to run the system. If every Player is required to possess the same environment dependencies, in order to run the game, additional friction is introduced to the proposed platform's "barrier-to-entry" for would-be Players. In addition, there is significant computational overhead of which is introduced when using a blockchain. While peer-to-peer capabilities will be exhibited, the computational overhead will exceed that of which consumer grade computers can handle.

The disadvantages of a decentralized architecture also include the inherent validation computations that are required to take place on a given Realm server. This leads to a performance issue with higher amounts of concurrent Players, beyond two Player Games.

\section{Guaranteed Risk Level}
This type of Game environment can easily be dominated by top cyber security professionals, opposed to new participants to the field. With this in mind, we propose a recurring reintroduction of "equi-value" risk into the Game mechanics. Since the Units are randomized, we propose an additional Game mechanic centered around the organized decommissioning of Player Units, to shuffle a new Unit in place of an old. 

Philosophically, if a defensive focused Player patches all of their Units, none of their Units can be exploited as easy as before. Depending on the Unit details, it may not be exploitable at all besides the network configurations denoted by the Unit at birth. Reintroducing risk such as this, enables ongoing work to be done by Players, even after securing an asset. A goal for a Player then becomes to remain adaptively hardened, not just for the Player to harden their assets. The leveling of the competitive landscape is conducive for proper gamification, and adoption.

\section{User-Computer Interface}
\begin{figure}[ht]
\centering
	\includegraphics[width=0.45\textwidth]{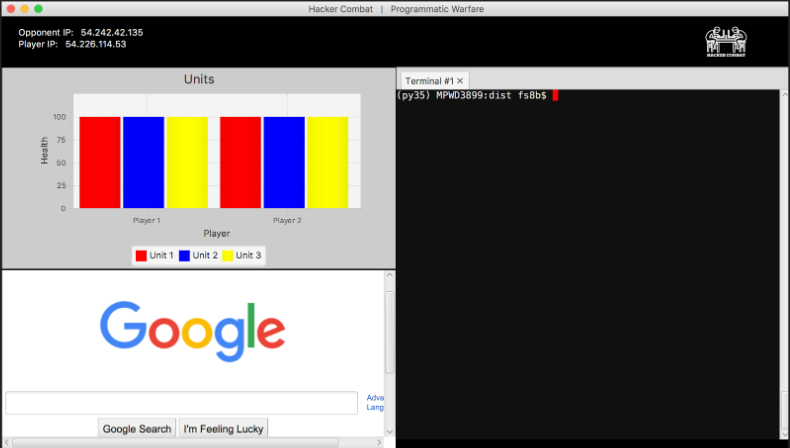}
	\caption{Demonstrating the GUI exposed to each player in the trials containing tabbed terminal window, one scoreboard, and one internet browser.}
	\label{utcinterface}
\end{figure}

We develop a bundled software package example that enables real-time monitoring and scoring by capturing Player typing data, Player system data, and other behavioral metrics. This enclosed environment allows for intricate an complex scoring mechanisms. The tradeoff is requiring users to obtain said bundle in order to participate. Figure \ref{utcinterface} demonstrates the experimental interface. In this section, we cover the segments of the interface, but do not dogmatically require it to exist. It is introduced as a mere convenience for the research.

\subsection{Scoreboard}
\ii
Throughout a competition, a player has direct access to real-time, Game Server centered scores on their Game Units. These metrics give universal Player metrics, from the "source of truth", the Game Server.

\subsection{Command Terminal}
\ii
Experiment participants are suggested to work from the built-in command terminal. This enables a full tracking of Player behavior. If a Player does not work within the confines of the built-in command terminal, the Player's actions, and behaviors are completely ambiguous to the Game system. One could require that all work be done from the proposed GUI, but development is needed to provide the same, holistic value as readily available platforms. One of which is a penetration testing framework known as Kali Linux. In the experimentation, we run the client on a Kali Linux instance as well. For the experiments, we monitor Player behavior within the proposed GUI, and restrict work outside of it. However, we ensure the proper usage of many industry penetration testing tools.

\subsection{Built-in Browser}
\ii
The browser contained within the bundle allows Players to reference internet resources from within the experimental bundle. This is useful for Players needing immediate answers to questions arising during gameplay. Users can navigate to, and from arbitrary internet resources; we model the client to run on both windows, and linux.

\section{Artificial Intelligence}

An additional motivation for this game is centered around the curiosity of creating an artificial agent that is capable of acting as a \textbf{dynamic} hacker. There are efforts from DARPA, on using Artificial Intelligence to train a system to hack another. However, the method used to train those systems does not originate from actual black hat actors, and their specific behaviors in given scenarios; current methodologies involve much "hard-coding" into the agent. We use AI to demonstrate how capturing cyber security tasks, command by command, can assist in the training of an artificial agent. We also show that upon taking a naive approach, a suitable agent can nonetheless be constructed. This is due to the tractability of the cyber security landscape and it is modeled here.

\begin{figure*}[ht]
\centering
	\includegraphics[width=\textwidth]{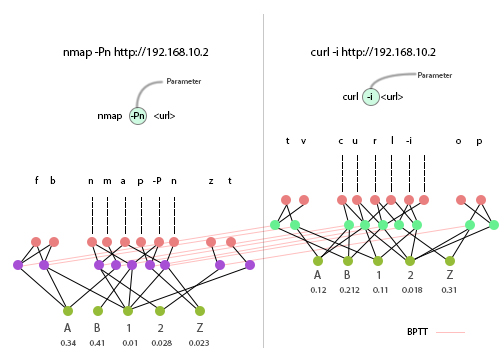}
	\caption{Illustrating the character-level RNN using back propagation through time (BPTT) for continuous next-character prediction. }
	\label{ainn}
\end{figure*}

We take two approaches to generate an artificial agent capable of competition. The two approaches use, Markov Decision Processes (MDP) for probabilistic reasoning about a finite, predefined Game environment, and a recurrent neural network (RNN) for character-level generation during each time step of gameplay. We also experiment with a hybrid approach of the two.

\subsection{Decision Making}

\begin{figure}[H]
\centering
	\includegraphics[width=0.5\textwidth]{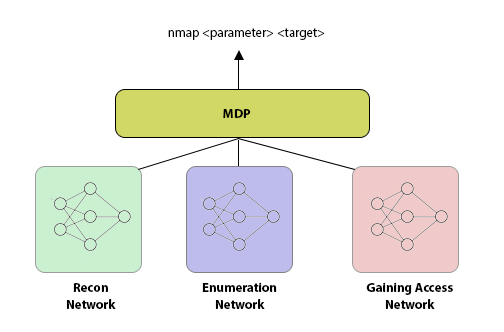}
	\caption{The decision making architecture for the proposed agent}
	\label{aiworld}
\end{figure}

\ii
To handle situational awareness, we model a probabilistic approach such as a Markov Decision Process (MDP), and a deterministic approach such as a Finite State Machine (FSM). Either way, this enables an agent to reason about some predefined world. The world of a hacker is a large, and diverse world, but we submit that it is not only a finite world, but is turing decidable, and tractable. Figure \ref{aiworld} shows the decision making architecture using three policy networks, and a Markov Decision Process for command orchestration.

\subsubsection{Markov Decision Process}
\ii
The first approach taken uses a Markov Decision Process (MDP) for probabilistic reasoning in the realm of common commands used in each phase of an attack. We realize this can only be as advanced as the known commands, suffering in autonomous creativity. This pure MDP approach is also limited by human cognition, and creativity. 

For a pure MDP implementation, we identify common commands in each phase of an attack. The automaton selects among possible phases, from its current phase. The Markovian property of the system requires that a decision at any given point in time can only rely on information known at its current state. 

\begin{equation}
	 p(s_{t+1} \mid s_t, a_t) 
\end{equation}

The transition probabilities describe the dynamics of the world. They play the role of the next-state function in a problem-solving search. Except that every state is thought to be a possible consequence of taking an action in a state. So, we specify, for each state \(s_t\) and action \(a_t\), the probability that the next state will be \(s_{t+1}\). You can think of this as being represented as a set of matrixes, one for each action. Each matrix is square, indexed in both dimensions by states. If you sum over all states \(s_t\), then the sum of the probabilities that \(s_t\) is the next state, given a particular state and action is 1 \cite{mdp}.

Once a transition phase is accomplished, the agent selects from the predefined set of suitable commands in its current phase. This approach is naive, but is suitable at progressing through a Game round. This is due to the finite nature of permutations for commands that exist in each phase. The first artificial agent approach only uses this MDP configuration. We also use an MDP to transition between phases for our Long-Short Term Memory (LSTM) RNN implementation. This hybrid approach is one of many possible configurations, and is chosen for the sake of this work alone.

\begin{equation}
	\sum p(s_{t+1} \mid s_t, a_t ) = 1
\end{equation}

We show that the sum of the probability of every available state, at a given time step \(t\), should equal 1.

\begin{equation}
	|S| \leq 5
\end{equation}

In one approach, we encode a finite set of commands per each hacker phase. We enable the agent to decide which phase to enter next, and to select among a finite set of commands for its selected time step phase. This approach is only as useful as the programming flexibility developed into it. It relies on human intuition to map suitable commands, in a predefined ecosystem to preferable world-space outcomes. Another problem is the requirement of anyone configuring this MDP, to have knowledge of state transition probabilities in an a priori manner.

\begin{figure}[H]	
\centering
	\includegraphics[width=0.45\textwidth]{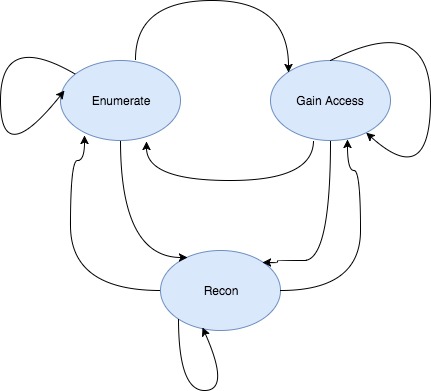}
	\caption{The FSM structure of the agent using an MDP}
	\label{aifsm}
\end{figure}

In another approach, the cardinality of the set of phases available at each step is equal to the number of total phases used to model the stages of an attack. We create a FSM by modeling hacker stages, recon, enumeration, and gaining access shown in figure \ref{aifsm}. It is possible to model every stage of any attack, but this requires more development time for encoding actions in each phase. Using an MDP for the purpose of competitive play, by itself, is a naive approach at best. For this matter, we also propose an approach inspired by Artificial General Intelligence (AGI). The AGI approach, also uses an MDP for phase state decision making.

The following pseudocode explains the general process for the proposed agent. We abstract the processes within planning the next state, and calculating the next command for the agent. Previous characters entered can be considered when performing prediction. Macro-level commands can also be represented as classes, and used for prediction.

\begin{algorithm}[H]
  \caption{Calculating the next agent state $s$, at time step $t$. We reference the MDP first, then predict using the relevant policy network. The ellipsis conveys the flexibility of using either character-level, or command-level prediction}
  \begin{algorithmic}[1]
    \Statex
    \Function{Decide}{$s_t$, $a_t$}
     \State $S \gets [\forall states \mid {s,t\ }] $ \Comment{$x$: state object array}
     \State $state_{chosen} \gets \argmax PlanNextState(S)$ \\
      \ \ \ \ \Return{$PolicyPredict(state_{chosen}, \ldots) $}
    \EndFunction
  \end{algorithmic}
\end{algorithm}

\subsubsection{Recurrent Neural Network using BPTT}
\ii
We initiate an agent purposed with generating terminal commands, at a character level. In one implementation, the agent autonomously navigates through each phase of an attack using an MDP. Each phase is assigned an individual policy network, built using an artificial neural network, of which is strictly trained on common commands used during that respective phase of an attack.

Using Back Propagation Through Time (BPTT) \cite{backprop}, each training iteration for the agent considers the past impressions of past inputs. Weights are considered between time steps. We do not take into account the previous command in time. For each policy network, the output is modeled stochastically, for each available character that can be used in a command. Each time step takes into account the previous letters typed, in the current command, and the agent's current state \(s_t\), or the current hacker phase.

When deciding the first character of a new command, Game metrics are taken into account. We do not propose a formal method of using Game metrics at this point, but we monitor the current phase in which the agent currently exists. We also model a knowledge-base inspired by the "5 W's": who, what, when, where, and why. If an agent doesn't known "who" the target is, the Recon phase is probabilistically preferred. If "who" is known, but "what" the target is, is not known, than the enumeration phase is probabilistically preferred, and so on. This is an arbitrary, and naive approach towards a Game specific knowledge base. However, it illustrates the capability of formulating an example agent. These factors enable for a tractable world view for the Markovian agent.

\subsection{Kali Linux}
\ii
Using Kali Linux, we utilize many tools to initiate attacks on target machines. The following is one example of a common command used in the "enumeration" phase of an attack.

\begin{verbatim}
      nmap <parameters> <ip or domain>
\end{verbatim}

In the "gaining access" phase, an agent can blindly fire every exploit available in Metasploit, or craft custom packets. Both approaches used warrant varied uses of metasploit at different times. For the MDP, we construct a finite state machine (FSM) using common commands in Kali Linux. The commands are categorized, by their purpose, into their respective hacker phases.

For character-level generation, training occurs with Kali Linux specific commands only. For the hybrid approach using RNNs, we create a FSM purposed with switching between modes, representing phases of an attack. We generate policy networks, of which are trained with commands specific to each phase. The agent reasoning within each phase is then generated on a character level.

\subsection{Simulation Training}
\ii
Training only occurs for the RNN approach. We utilize human subjects to train each policy network with the phases of an attack chosen. Strictly, subjects input penetration test commands used for the respective phase undergoing training. This paper does not highlight the variety among commands used to train the networks, in each phase. However, the usage of Kali Linux's toolset enables a high variety of command availability.

\subsection{Data Capture}
\ii
In equation \ref{eq:cmd_time}, we show how to calculate a users effective CoPM. We can use the same capturing tasks from the system to generate other metrics. We create a knowledge base consisting of the command that a Player commits, the scenario in which it is committed, and the outcome of the Player action. We also experiment with an agent that does not receive any feedback beyond that of knowing whether or not a used command is \textbf{valid}. The agent that does not receive feedback is solely purposed with generating character-level predictions based on previous characters used, and the phase with which it exists at a given time \(t\).

\subsubsection{Command Capturing}
\ii
One of the benefits of providing a work environment for Players is the ease with which command capturing can be accomplished. Every time a Player presses "enter" in their terminal window, the characters they type are captured, with their ordering recorded. We also capture character level data. Upon the pressing of a key, the system logs it. We then reconstruct commands when needed. This is also how EPM is monitored, as covered before. If a user presses any special representation of entering a command, it is taken into account. This is similar to pressing an "up" arrow, or "!!", both to refer to the last command entered. The following is the command shown in figure \ref{aifsm}. It is used to enumerate a target machine.

\begin{verbatim}
        nmap -Pn 192.168.10.2
\end{verbatim}

If we calculate the hamming distance between the previous command, and the current, we can monitor differences in commands. This is not formally included in any experimentation, but enables an additional perspective on gameplay per each command. The following is a snippet to calculate the hamming distance between two commands, which are just of the standard String data type: 

\begin{verbatim}
  def hamming_distance(c1, c2):
    if len(c1) != len(c2):
      raise ValueError("Unequal length")
    return sum(el1 != el2 for el1, 
                      el2 in zip(c1, c2))
\end{verbatim}

\subsection{Feature Selection}
\ii
We model our feature selection efforts to account for each letter that can be inputted into a terminal window. Representing each possible character as a class. We abstract away the use of "targets", such as IP Addresses; these are inputted in whole using predefined flags. For character-level generation, this introduces a problem of knowing the accurate parameters that belong to commands. To input a target IP Address during a generated command is another problem to be solved. It is one problem to learn terminal commands by character, but another to do so by abstracting away the use of parameters.

To abstract the problem away, we assign a binary class to represent "IP Address". As a data structure, this feature represents the occurrence of an IP Address. This ensures that an agent can ignore all characters indicating an IP Address, or variable target. The agent can then keep track of whether or not it should insert the target IP Address as a "next character".

\subsection{Policy Networks}
\ii
We generate policy networks for each stage in the phases of an attack covered in the previous Gameplay section. This gives us a separate neural network architecture for each phase. Each policy network maps previous character input, to next character input. We combine this with an MDP to handle the decision of the next state selection. While in a given state, the policy network outputs the probability for each possible character, of which the most likely character in selected as "next".

This process continues until a command should be entered. We also generate a policy network for determining the "end" of a proposed command that checks for the "validity" of a submitted command. Variable information, and parameters in each command are not used during training. This translates to the removal of all domain, and IP addresses from each command. This is the case if the Game environment is fully observable. The set of all possible characters is denoted by \(CH\) 

\begin{equation}
	\{ CH  \mid A, B, 0, 9, -, !, \ldots \ ch_n \} 
\end{equation}

The sum of all of the probabilities of the next character is equal to 1. The agent probabilistically considers all possible characters,

\begin{equation}
	c_{ch} \in CH, \sum_{ch=1}^{|\ CH\ |} p(c_{ch}) = 1
\end{equation}

We can then interchange one policy network's output for another's input. This is subject for an entire paper, but we make note of combining the previous output from a "Recon" stage command, into the input for a "Gaining Access" phase computation. This is an attempt to add contextual awareness to the LSTM-RNN process. We use Tensor Flow \cite{tf} for all AI work.

\subsubsection{Recon Network}
\ii
The Recon policy network is trained on Recon specific tasks. Among these tasks include scraping target web applications, and accessing publicly available information. Even though "Google" scraping is developed into the proposed agent, this is not useful during matches against a regular Player. There is no information that an agent can discover on the public internet that is conducive to victory against a conventional opponent. We train the Recon Policy network on commands from platforms like "Curl", and "Scrapy". This enables a network for pure Recon tasks. Although there is a thin line between Recon, and Enumeration, we purposely delineate the two for this work.

\subsubsection{Enumeration Network}
\ii
Following the pattern, the Enumeration Policy network is trained on commands from platforms like "nmap", "netcat", "hping", and other enumeration tools that exist within Kali Linux. The Enumeration network can retrieve information of which the Recon, and Access network commands cannot. Due to each network being trained independently, this is true for each network.

\subsubsection{Access Network}
\ii
The last Policy network trained is that for Gaining Access. There are many tools that exist in Kali Linux to accomplish this, and we train the network on all of them that are available. We also enable the capability for the agent to probabilistically launch every attack technique known to Kali Linux. This can be thought of as a "hail mary" approach, and is used as a last resort.

\subsection{Outcome Analysis}
\ii
An agent can learn on every command upon command submission, or at the end of Game. In both cases the Agent creates a map between the characters used in the command, and an outcome. The two approaches differ in outcome used. For this work, we use a combination of monitoring the Game environment differently, including the "health" of Game Units for character-level validity checks.

\subsubsection{Command Outcome Analysis}
\ii
If an agent is to learn on each command, the outcome to map can be the health of the Player's Units. However, this implementation is too ambiguous for learnability. We examine further metrics from the environment of which to use during modeling.

\subsubsection{Game Outcome Analysis}
\ii
To learn on the outcome of a Game, the agent creates a map between all commands used during the Game, and whether or not the Player "Wins". This yields a binary outcome, of which we represent as the set \(\{0,1\}\). If the agent loses a given Game, it will penalize every action taken during the Game by the penalty rate, \(\alpha\). This can skew correct actions in the wrong direction, but the assumption is that with enough rounds of gameplay, on average, the correct decision can be made easier by the agent.

\section{Physical Apparatus Extension}
This project implies that such a Game system can be combined with real-world, physical actuators in the service of creating an additional layer of entertainment around the core Game concepts. It is not the aim of this paper to identify such configurations. However, it is implied that the proposed system enables a physical augmentation. The use of physical augmentation of these Game mechanics causes the system to resemble a modern TV Game Show program.

\section{Further work}
One evident factor of this work is the strict physiological testing on male gendered participants. We look to iterate upon this work using a female audience. Another factor upon which we would like to iterate is the type of scoring mechanism around exploiting a system. We only take into account exploits that enable remote code execution. Since we model Units with a property known as "health", it skews the type of scoring mechanism quite a bit. Without the notion of "health", one could even more closely examine the characteristics of XSS, or SQL Injection. These may require more dependencies for the system, and more development of specific scoring capabilities. The vast landscape of vulnerability types, and exploitation types warrant more research in the realm of suitable scoring mechanisms.

We also propose more work to be done in the realm of Artificial Intelligence. Though the proposed approach is reliable, there exists a plethora of techniques that can be used for this work. As AI evolves over time, this should be a continuous topic of discussion, as this work brings science closer to AGI capable of penetration testing a myriad of diverse systems.

Furher work is warranted to visualize Player physiological metrics in real-time. It is understood that many of the methods used in this paper provide challenge towards this goal. 

\section{Conclusion}
In conclusion, this article introduces a new paradigm of competitive, head-to-head cyber warfare. We demonstrate a real-time system capable of handling the demands of such a system. Player performance is also monitored during game play by analyzing Player efficiency, and physiological metrics. Lastly, we demonstrate multiple approaches in the pursuit of developing an artificial agent capable of participating in the proposed framework, but playing at a competitive level based on previously recorded gameplay rounds.

\end{document}